\documentstyle[preprint,aps]{revtex}

\begin{document}
\tighten
\draft
\title{Weak reactions on $^{12}$C within the Continuum Random Phase
Approximation with partial occupancies}
\author{E.~Kolbe$^1$,  K.~Langanke$^2$ and P. Vogel$^3$}
\address{
$^1$ Institut f\"ur Physik, Universit\"at Basel, Basel, Switzerland \\
$^2$ Institute for Physics and Astronomy, 
Aarhus, Denmark \\
$^3$ Physics Department, California Institute of Technology, Pasadena,
California 91125, USA}

\maketitle

\begin{abstract}
We extend our previous studies of the neutrino-induced
reactions on $^{12}$C and muon capture to include partial 
occupation of nuclear subshells in the framework of the
continuum random phase approximation.
We find, in contrast to the work
by Auerbach et al., that a partial occupation of the $p_{1/2}$
subshell reduces the inclusive cross sections only slightly. 
The extended model
describes the muon capture rate and the $^{12}$C($\nu_e,e^-$)$^{12}N$ cross
section very well. The recently updated flux and the improved model
bring the calculated $^{12}$C($\nu_\mu,\mu^-$)$^{12}$N cross section 
($\approx 17.5 \cdot 10^{-40}$ cm$^2$) and the data
($(12.4 \pm 0.3({\rm stat.}) \pm 1.8({\rm syst.)} \cdot 10^{-40}$ 
cm$^2$) closer together, 
but does not remove the discrepancy fully.
\end{abstract}


\section{Introduction}

In recent years the study of neutrino-nucleus reactions has attracted
continuously growing interest. Besides astrophysical motivations 
this interest has particularly been inspired by the use of
neutrino-nucleus reactions as a laboratory to study fundamental
properties of neutrinos. 
The best studied nucleus by $\nu$-induced
reactions is
$^{12}$C, which is an ingredient of liquid
scintillator and
plays an important role in the oscillation search experiments performed by the
KARMEN \cite{Drexlin} and LSND \cite{Albert} collaborations.

The exclusive and inclusive charged-current reactions
$^{12}$C($\nu_e,e^-$)$^{12}$N have been measured by the KARMEN and LSND
collaborations for neutrinos generated by muon decay at rest
\cite{Drexlin,Albert,Kleinfeller,Imlay}. The KARMEN
collaboration has supplemented this by a measurement of the
neutral current
$^{12}$C($\nu,\nu'$)$^{12}{\rm C}^*$(15.11 MeV) cross section \cite{neutral}.
Using the LAMPF
pion-decay-in-flight neutrino source, the LSND collaboration has also
measured the $^{12}$C($\nu_\mu,\mu^-$)$^{12}$N cross section \cite{LSND}. 
Together with the precisely known inclusive and exclusive
muon-capture rates on $^{12}$C \cite{muon}, 
these measurements represent a 
challenge to the theorists.

Various data on the exclusive reactions (also
including electro-magnetic transitions and beta decays), leading to the
triad of $T=1, J^{\pi}=1^+$ 
analog states in the $A=12$ nuclei (these are the $^{12}$B
and $^{12}$N ground states and the excited state in $^{12}$C at 15.11
MeV), appear to be consistent and are theoretically
well described, validating the total neutrino
flux determination \cite{Engel}.
However, the situation is somewhat different for the inclusive reactions;
the term is  adopted here for the reactions leading to final states in
$^{12}$B and $^{12}$N other than the ground state 
(in the following denoted by $^{12}$B$^*$ and $^{12}$N$^*$, respectively) 
and involving 
states in the continuum.
The data on inclusive weak processes 
($^{12}$C($\nu_e,e^-$)$^{12}$N$^*$, $^{12}$C($\nu_\mu,\mu^-$)$^{12}$N$^*$ 
and muon capture)
are usually not simultaneously reproduced by calculations 
\cite{Kolbe94,Kolbe95}, 
with a notable exception \cite{Auerbach}. 
For example, a reasonable model to calculate these cross sections is the
continuum random phase approximation (RPA). In fact, one finds that
this model describes the inclusive muon capture rate and the
$^{12}$C($\nu_e,e^-$)$^{12}$N$^*$ cross section very well, but it
overestimates the 
$^{12}$C($\nu_\mu,\mu^-$)$^{12}$N$^*$  data noticeably.
Of course, these three weak processes probe the nuclear response at different
momentum ($q$) and energy ($\omega$) transfers, and thus the good
description of the capture rate
(with average values 
 $\overline{\omega} \approx 25$ MeV, $\overline{q} \approx 80$ MeV/c)
and the $\nu_e$-induced cross section 
($\overline{\omega} \approx 23$ MeV, $\overline{q} \approx 50$ MeV/c)
might simply indicate that the model does well for low values of $q$ and
$\omega$, but it is less reliable at larger momentum and energy
transfers which are important for the 
$^{12}$C($\nu_\mu,\mu^-$)$^{12}$N$^*$ experiment
($\overline{\omega} \approx 37$ MeV, 
$\overline{q} \approx 200$ MeV).  
Such hypothesis has, however, been invalidated by the 
continuum RPA study of the
inelastic electron scattering on $^{12}$C where the model has
been demonstrated to describe well the data for momentum and energy
transfer regions relevant for the LSND experiment \cite{Kolbe97}.

Could the origin of the discrepancy be nuclear correlations missing in
the continuum RPA model? This is in fact 
what the authors of Ref. \cite{Auerbach} suggest.
First, they point out that, due to pairing effects, there are
about 1.6 nucleons occupying the $p_{1/2}$ subshell in the $^{12}$C ground
state, rather than the closed $p_{3/2}$ configuration (and empty
 $p_{1/2}$ subshell) assumed in the previous
continuum RPA calculations. Second, they argue on the basis
of a standard RPA
calculation that this partial occupation in fact allows one to
simultaneously describe all three inclusive weak processes. 

Obviously the method of
choice to test this conjecture is the interacting shell model. But a
reliable calculation of the
$^{12}$C($\nu_\mu,\mu^-$)$^{12}$N$^*$ cross section requires inclusion
of at least the 
complete 3$\hbar\omega$ model space, which is a formidable task
even on modern computers. 
Short of such a large-scale shell model
calculation, we have performed a set of 
continuum (standard) RPA calculations for the
various inclusive (and exclusive) cross sections extending our previous
model to allow for partial occupancies of the subshells. 
The second
motivation for repeating our previous calculations within an improved
model is that recently the decay-in-flight
neutrino spectrum for the LSND experiment has
been revised. With the updated flux, the experimental cross section
slightly increased to 
$(12.4 \pm 0.3({\rm stat.}) \pm 1.8({\rm syst.)} \cdot 10^{-40}$ 
cm$^2$ \cite{Imlay2}.
As the LSND
experiment is such an important benchmark we feel obliged to also
consistently update our calculated cross section allowing for a
meaningful comparison.

The continuum random phase approximation for nuclei with closed-shell
configurations  is well documented in the literature
\cite{Kolbe92}. A generalization of this formalism to partial
occupancies is readily achieved by multiplying the relevant matrix
elements in Eq. (17) of Ref. \cite{Kolbe92} 
with the  partial occupation numbers, $n_h$, for the hole
states, thus
\begin{equation}
\langle {\rm p h^{-1}} | V | {\rm p' h'^{-1}} \rangle
\rightarrow
n_h n_{h'} \langle {\rm p h^{-1}} | V | {\rm p' h'^{-1}} \rangle.
\end{equation}
After this replacement, the numerical solution of the Continuum RPA
equations follow the lines as outlined in Ref. \cite{Kolbe92}. The
orthogonalization of the particle states on the hole states is performed
by Schmidt's procedure. 

With the exception of the $T=1$ triad of bound states
in the $A=12$ nuclei,
all other final states (which contribute to the inclusive cross section)  
lie above the particle emission threshold and
therefore are described as resonances within  the continuum RPA model.
As residual interactions we used as before the finite-range G-matrix \cite{Bonn} 
derived from the Bonn NN potential (BP) and, in order to estimate 
the theoretical uncertainties, also an empirical zero-range 
Landau-Migdal force (LM).

To derive an estimate of the
partial occupancies of the $p_{1/2}$ subshell, we have performed a shell
model calculation for $^{12}$C with the OXBASH code \cite{Oxbash}
and the Cohen-Kurath
interaction \cite{Cohen}. 
This calculation suggests a partial occupation for the
$p_{1/2}$ subshell of 0.75 for protons and neutron, respectively. 
This result is rather close to the value 0.8 used in Ref. \cite{Auerbach}.

First we recalculated the various exclusive processes leading to the
$T=1$ triad in the $A=12$ nuclei. Since these
states are discrete we used the standard RPA
in this calculation, again including partial occupancies as outlined
above. As single particle states we considered the complete
$(1s)(1p)(2s1d)(2p1f)$ model space. For the proton and neutron $p_{3/2}$ and
$p_{1/2}$ single particle energies we used the experimental values; the
others have been determined from a Woods-Saxon potential which describes
the $^{12}$C charge distribution well \cite{Co85}. We have checked that an
enlargement of the model space does not affect our results.
Our RPA results are compared to data in
Tables~1 and~2. 

In Table ~1 the rescaling factors $N$ needed to bring calculation and data in
agreement are shown. Note, that allowing the partial occupancy of the 
$p_{1/2}$ subshell reduces the value of $N$ substantially, from about
4 to less than 2. At the same time
the rescaling factors are essentially identical
for the three processes considered, showing consistency.
The necessity of using such renormalization 
reflects again the well known fact that random phase
approximation calculations do not describe all (proton-neutron)
correlations necessary to reproduce the quenching of the Gamow-Teller
strength \cite{Muto,Bertsch}.
This aim can be only achieved in shell-model
calculations within a complete major shell, introducing additionally the
renormalization of the spin operator by the universal factor $0.77$
\cite{Brown88,Langanke95,Martinez96}. Deviating from this universal
finding, the Cohen-Kurath interaction had been designed to describe the
weak interaction transition rates in $^{12}$C without incorporating
the renormalization of the spin operator. Thus the fact that $N>1$ in
our calculation 
again shows the wellknown result that
the reproduction of the Gamow-Teller strength requires the consideration
of genuine correlations beyond a simple mean-field approach.

In Table ~2 we compare  the results on the exclusive
neutrino induced reactions with data. The entries are based
on $N^{new} = 1.5$ and $N^{old} = 4.0$.
The agreement remains good,
and at the same time, as mentioned above,
 the required renormalization is 
 substantially smaller than in our
previous RPA calculation.  
The reduction is caused by the
destructive interference of the $p_{3/2}$ and $p_{1/2}$ configurations
in the RPA description of the Gamow-Teller transition, as discussed by
Auerbach et al. \cite{Auerbach}. 

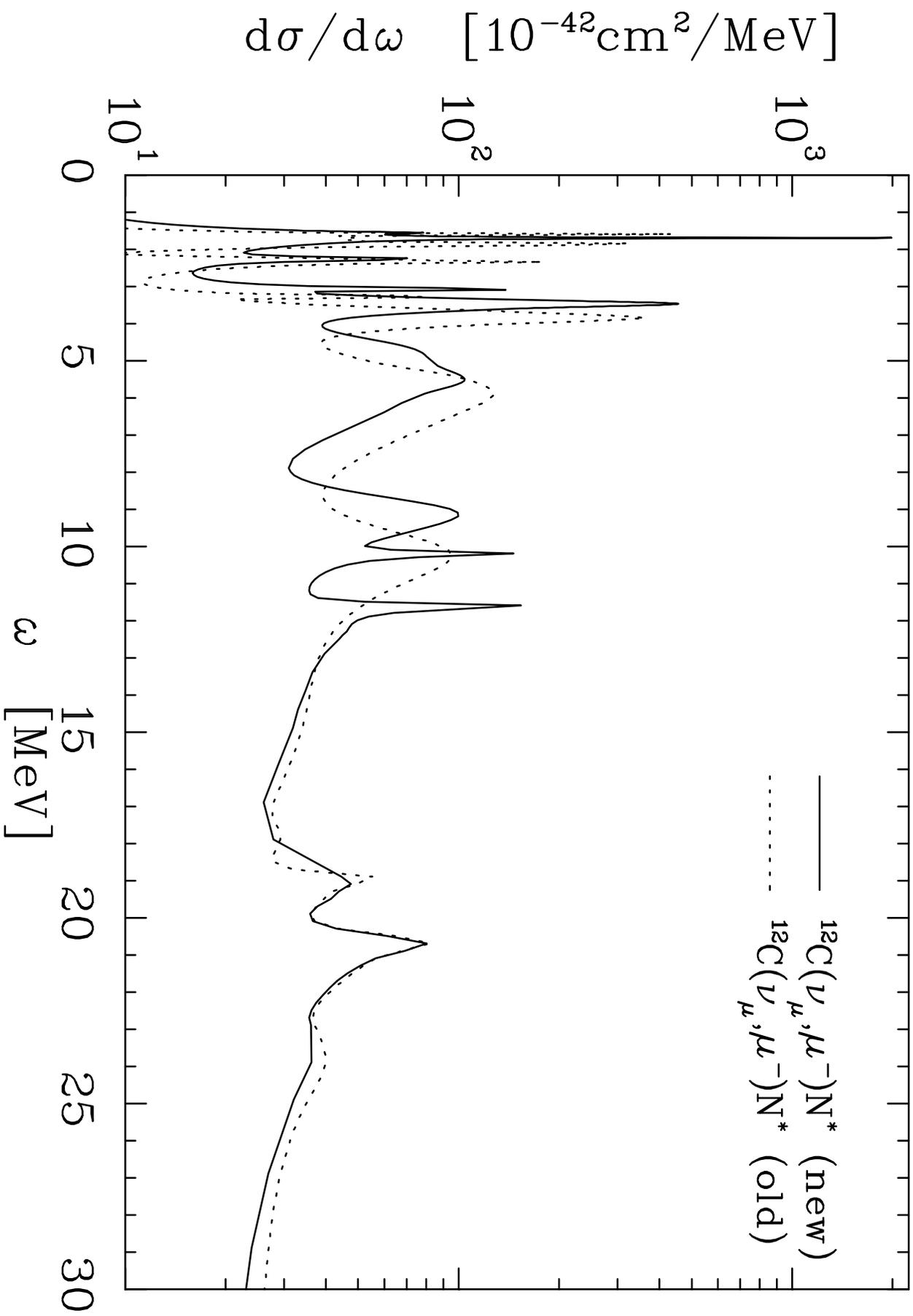
\begin{figure}
  \caption{The $^{12}$C($\nu_\mu,\mu^-$)$^{12}$N$^*$ cross sections as a 
           function of the excitation energy $\omega$ of the nucleus.
           The present results obtained with partial occupancy of the
$p_{1/2}$ orbital (labelled `new') are compared to those assuming a pure $(p_{3/2})^8$
closed-shell configuration for the $^{12}$C ground state (labelled `old').}
\label{Fig1}
\end{figure}

We  then studied the three inclusive processes within our improved
continuum RPA model. 
The noticeable mixture of $p_{1/2}$ configuration into the
$^{12}$C ground state allows for the excitation of a larger number of
resonances than in the previous calculation. 
This is demonstrated in Fig.~\ref{Fig1}, where the 
$^{12}$C($\nu_\mu,\mu^-$)$^{12}$N$^*$ cross section 
(calculated with the BP force) is shown as a function of the excitation 
energy $\omega$ of the nucleus.
However,  this figure also demonstrates that the strength going into these
additional resonances is taken away from those transitions already
present for the pure $p_{3/2}$ ground state. In fact, Table~3 shows
that the
total inclusive rate and cross sections are only slightly reduced
by the mixing of a $p_{1/2}$ component into the ground state. 
The inclusive muon capture rate and the 
$^{12}$C($\nu_e,e^-$)$^{12}$N$^*$ cross section 
agree very well with data: 
For muon capture the improved continuum RPA yields for both residual
interactions rates close to
the experimental value ($32.8 \cdot 10^3$ s$^{-1}$), and the
calculated
$^{12}$C($\nu_e,e^-$)$^{12}$N$^*$ cross sections are 
also close to the most recent data $(5.1 \pm 0.6 \pm 0.5) \cdot 10^{-42}$ cm$^2$
\cite{Kleinfeller} and
$(5.7 \pm 0.6 \pm 0.6) \cdot 10^{-42}$ cm$^2$ \cite{Imlay}.
Our calculation of the inclusive
$^{12}$C($\nu_\mu,\mu^-$)$^{12}$N$^*$ cross section 
results also in a reduced value
compared to our previous study, where the reduction is about half 
($0.8 \cdot 10^{-40}$ cm$^2$ for BP, 
$ 1.9 \cdot 10^{-40}$ cm$^2$ for LM) due to the revised flux and half 
($0.7 \cdot 10^{-40}$ cm$^2$ for BP,
$0.9 \cdot 10^{-40}$ cm$^2$ for LM) due to the partial 
occupation of the $p_{1/2}$ subshell in the ground state wave function.
Our new total 
$^{12}$C($\nu_\mu,\mu^-$)$^{12}$N cross section (inclusive + normalized
exclusive) is then 
$17.8 \cdot 10^{-40}$ cm$^2$ and
$17.5 \cdot 10^{-40}$ cm$^2$ for the BP- and LM-force, respectively, 
to be compared with the last experimental value of 
($(12.4 \pm 0.3({\rm stat.}) \pm1.8({\rm syst.)} \cdot 10^{-40}$ 
cm$^2$). 
Thus the disturbing discrepancy is less severe than before, 
but it not entirely removed.

Our result does not support in detail the argumentation in Ref.
\cite{Auerbach} that the discrepancy between RPA calculations and data
can be essentially fully  removed if $p_{1/2}$ partial occupancy is considered (Ref.
\cite{Auerbach} gives $(13.5-15.2) \cdot 10^{-40}$ cm$^2$) and all inclusive
and exclusive processes can be simultaneously and consistently
described. We also mention that Singh et al. \cite{Singh} have studied
the various processes based on the local density approximation and find
a total (exclusive + inclusive) muon capture rate of 
$(3.6 \pm 0.22) \cdot 10^4$ s$^{-1}$ \cite{Mukho} (the
experimental value for the total rate is 
$(3.8 \pm 0.10) \cdot 10^4$ s$^{-1}$) and a
$^{12}$C($\nu_\mu,\mu^-$)$^{12}$N cross section of 
($16.5 \pm 1.3) 10^{-40}$ cm$^2$ \cite{Singh}, slightly lower, 
but consistent with
our new result. Notably, Mintz and collaborators predicted the
$^{12}$C($\nu_\mu,\mu^-$)$^{12}$N cross section \cite{Mintz}, 
but the model assumptions
within the elementary particle model used in the approach of these
authors are questionable and not justified for this reaction
\cite{Kolbe97}.

In summary, we have found that the consideration of a partial $p_{1/2}$
occupancy lowers the exclusive cross sections and muon capture rate
noticeably, but it reduces  the results for the inclusive
processes only slightly. The reason for the reduction of the exclusive
cross sections has been discussed above. But can one 
understand why the continuum RPA is apparently able to describe the inclusive
processes without the need for parameter adjustment, unlike the
exclusive Gamow-Teller transitions? Another way of thinking about this
problem is the question to what extent the correlations of nucleons
within the $p$-shell influence the results. To shed the light on this, one
can evaluate the total strength of various operators ${\hat O}$, i.e.
the norm of the vector ${\hat O} | g.s. \rangle$, with and without the
effect of correlations. Note that the total strength depends only on the
ground state wave function and is therefore relatively easy to evaluate.
(Here we follow the points made in Ref. \cite{WEIN}).

For the positive parity operators this is done in Table~4.
Table~4 illustrates the well known fact that the $p$ shell correlations
are very important for the Gamow-Teller operator $\sigma \tau$. (Note
that the RPA gives the total Gamow-Teller strength of 5.5, only slightly
reduced when compared to the `naive' estimate of the independent
particle model. Also, the exact shell model predicts that a strength of
0.12 goes to excited $1^+$ states, which are absent in the naive model.)

But the situation with the quadrupole operators is rather different. The
total $p$ shell strength of the spin-independent operator, and the
strength summed over the multipoles of the spin-dependent operators
is affected by the correlations only on the $30-40 \%$ level, even
though the individual spin-dependent multipoles are affected more.
Moreover, the $p$ shell strength represents only a small fraction of the
total quadrupole strength, which is concentrated in the $2\hbar \omega$
excitations, unaffected by correlations as long as we assume that the
ground state has only $p$ shell nucleons.

The lowest $2^+$ resonance has a dominating $(0p)$ configuration and
thus its contribution to the various cross sections might 
be affected by the present more flexible description of the $^{12}$C
ground state. We find that the contribution 
of the $2^+$ multipole to the $\nu_e$-induced cross
section and to the muon capture rate increases slightly (by $0.02 \cdot
10^{-42}$ cm$^2$ and $0.17 \cdot 10^3$ s$^{-1}$, respectively), but is
still negligible compared to the total cross section and rate ($0.8 \%$
and $2.8 \%$, respectively). For the $\nu_\mu$-induced reaction, the
effects of the
partial occupancy formalism  and of the improved experimental neutrino
flux nearly cancel each other so that the partial cross section of the $2^+$
multipole slightly decreases (by $0.08 \cdot 10^{-40}$ cm$^2$). As the total
cross section is also reduced, the contribution of the $2^+$ multipole,
of which the $(0p)$ configuration is only a small fraction, to
the total cross section is
increased to $12 \%$ in our present calculation.

\begin{figure}
  \caption{Total strength of the dipole (upper part) and quadrupole
           (lower part) operators versus the occupation of the 
           $p_{1/2}$ subshell. 
           The curves are labeled by the corresponding angular momentum of
           the $r^l (Y_l \sigma)^{I=\lambda}$ operators for $l=1$ and $l=2$.}
  \label{Fig2}
\end{figure}
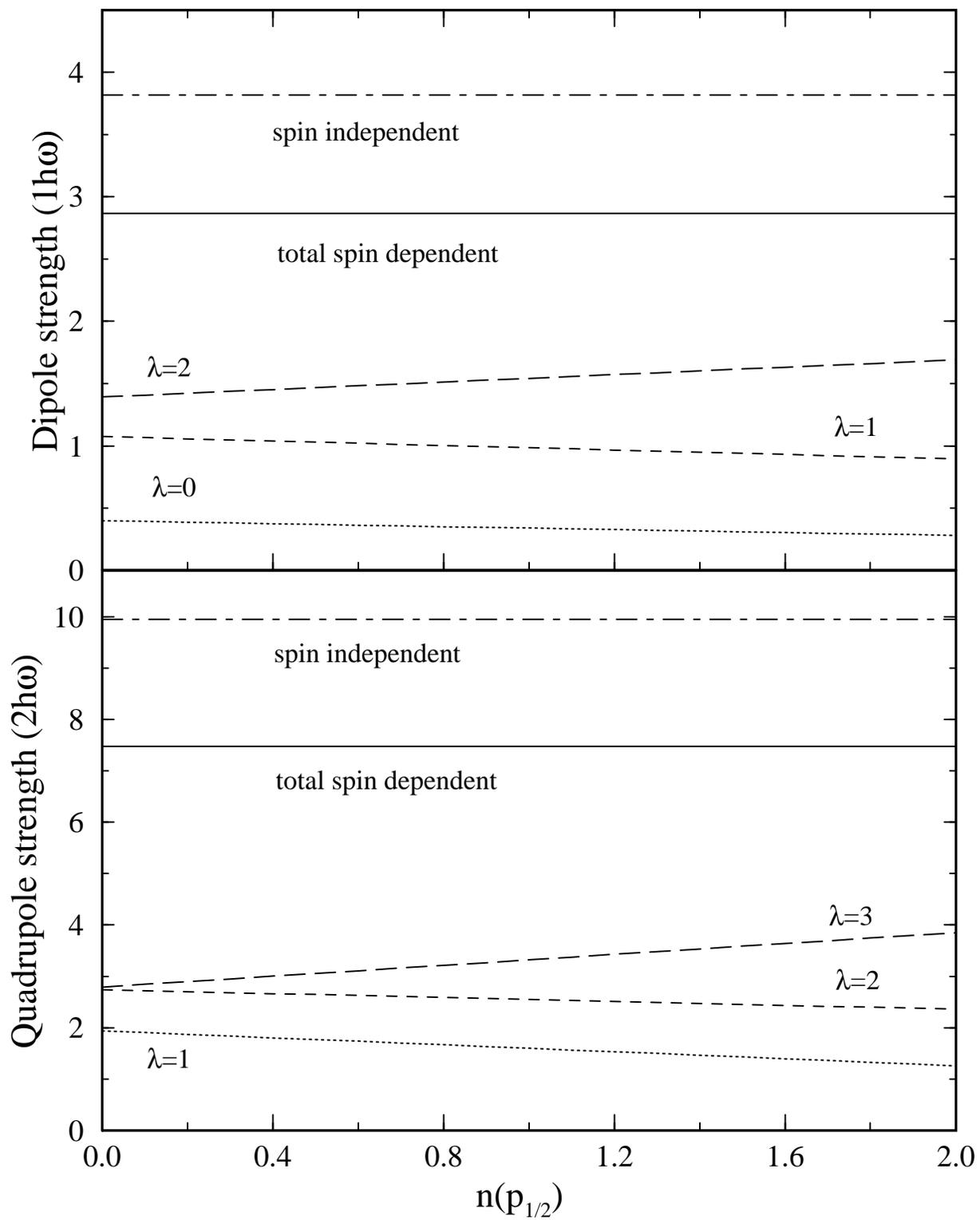

However, the inclusive reactions we are considering are dominated by the
excitations of the negative parity states. To what extent does the
strength depend on the occupation of the $p_{1/2}$ subshell for these
operators? According to Ref. \cite{Auerbach} the cross sections should be
noticeably reduced when there are about 1.6 nucleons in the $p_{1/2}$
subshell. In order to test this point, 
we plot in Fig.~\ref{Fig2} the dipole and quadrupole
strength as a function of the $p_{1/2}$ subshell occupation. We find
that when summed over multipoles the strength is totally independent of
this occupation number. But even the individual multipoles depend on the
occupation numbers only mildly. We find that the $p$ shell correlations
are relatively unimportant for the full strength.

Of course, the partial occupation of the $p_{1/2}$ subshell leads to the
occurrence of new states, reshifting the strength. As the $p_{1/2}$
orbital reflects an excited state compared to the $p_{3/2}$ orbital, the
strength is shifted slightly to higher energies.
Thus, even if the total strength is unchanged, this redistribution of
strength changes the cross section due to the related change of kinematics.
As the strength is shifted to higher energies, its weight in the cross
section is reduced, as apparent from our results. The relative
importance of this effect decreases with increasing
momentum and energy transfer, explaining why we find a reduction of
about $20 \%$ for the $\nu_e$ induced experiment, while it is only $4\%$
for the $\nu_\mu$-induced reaction.

The LSND collaboration has recently conducted a search for 
$\nu_\mu \rightarrow \nu_e$ oscillations for high energetic 
$\nu_\mu$ ($E_{\nu_\mu}\leq 300$~MeV) stemming from
$\pi^+$ decay in flight. 
The oscillation signal consists of isolated, high-energy electrons 
(60~MeV$\leq E_{e^-} \leq$200~MeV) in the detector coming from
charged current $^{12}$C($\nu_e,e^-$)$^{12}$N reactions.
Such events cannot be caused by $\nu_e$-neutrinos 
from pion decay at rest, which are also produced in the beam stop 
at LAMPF, because their energies is lower than $52.8$~MeV.
In the experiment an excess in the number of these events was found, 
which, if interpreted as an oscillation signal, leads to an
oscillation probability of $(2.6 \pm 1.0 \pm 0.5) \times 10^{-3}$~\cite{LSND2}.
This value and its systematic error depends dominantly on the knowledge
of the inclusive $^{12}$C($\nu_e,e^-$)$^{12}$N cross section 
and the absolute neutrino flux through the detector.
To aid the analysis, we present in Table~5 this cross section calculated within the
improved RPA model. 
Comparing with our previous results in Ref.~\cite{Kolbe95}, we find 
for most of the neutrino energies a reduction of the cross section
between 5--10\%. 
As this reduction is smaller than the 10\%-error attributed to the 
uncertainty of $^{12}$C($\nu_e,e^-$)$^{12}$N cross
section in the LSND-analysis~\cite{LSND2}, it does not
require a re-analysis of the experiment. 

{\it Acknowledgments}
We were supported in part by the Swiss National Science Foundation, the
Danish Research Council and by the U.S. Department of Energy under
grant DE-FG03-88ER-40397.

\begin{table}
\begin{center}
\caption
{The rescaling factors $N$ of the matrix elements for the exclusive 
 transitions were determined for both of the applied residual interactions
 by comparison of the calculated rates 
 ($\omega_{\rm BP}^{\rm new}$, $\omega_{\rm LM}^{\rm new}$) 
 with the experimental beta-decay rates and partial muon capture rate to 
 the $^{12}$B ground state (in units of $s^{-1}$).
 Also shown are the previous results 
 ($\omega_{\rm BP}^{\rm old}$, $\omega_{\rm LM}^{\rm old}$)
 obtained for a completely filled $p_{3/2}$ subshell.}  
\begin{tabular}{|c|cr|cccr|cccr|} \hline
   process & data & Ref. & $\omega_{\rm BP}^{\rm new}$ 
                         & $\omega_{\rm LM}^{\rm new}$ 
                         & $N^2_{\rm BP}$ & $N^2_{\rm LM}$ 
                         & $\omega_{\rm BP}^{\rm old}$ 
                         & $\omega_{\rm LM}^{\rm old}$
                         & $N^2_{\rm BP}$ & $N^2_{\rm LM}$
                         \\ \hline
   $^{12}$C($\mu^-,\nu_\mu$)$^{12}$B$_{\rm g.s.}$  
     & $6 \: 050 \pm300 (*) $   & \protect\cite{Bu70,Gi81,Mi72,Ro81} 
     & $  9 \: 330 $ & $  8 \: 538 $ & 1.54 & 1.41   
     & $ 22 \: 780 $ & $ 25 \: 400 $ & 3.77 & 4.20 \\
   ${}^{12}{\rm B}_{\rm g.s.} (\beta^-) {}^{12}{\rm C}_{\rm g.s.}$ 
     & $33.36 \pm 0.13$ & \protect\cite{Aj90} 
     &  49.6  &  44.3  & 1.49 & 1.33 
     & 123.6  & 128.8  & 3.71 & 3.86 \\
   ${}^{12}{\rm N}_{\rm g.s.} (\beta^+) {}^{12}{\rm C}_{\rm g.s.}$ 
     & $59.58 \pm 0.46$ & \protect\cite{Aj90} 
     &  98.7  &  88.2  & 1.66 & 1.48  
     & 247.1  & 257.4  & 4.15 & 4.32 \\
\end{tabular}
\end{center}
\end{table}

\begin{table}
\begin{center}
\caption
{The exclusive cross sections for 
 charged and neutral current neutrino scattering 
(in units of $10^{-42}$ cm$^2$).
The results of our improved RPA calculation 
for both of the applied residual interactions
($\sigma_{\rm BP}^{\rm new}$, $\sigma_{\rm LM}^{\rm new}$) 
are compared to the data and the previous RPA calculation 
($\sigma_{\rm BP}^{\rm old}$, $\sigma_{\rm LM}^{\rm old}$) 
without partial $p_{1/2}$ subshell occupation.}
\begin{tabular}{|c|cr|rr|rr|} \hline
   process & data & Ref. 
     & $\sigma_{\rm BP}^{\rm new}$ & $\sigma_{\rm LM}^{\rm new}$ 
     & $\sigma_{\rm BP}^{\rm old}$ & $\sigma_{\rm LM}^{\rm old}$ \\ \hline
   $^{12}$C($\nu_e,e^-$)$^{12}$N$_{\rm g.s.}$      
     & $10.5\pm1.0({\rm stat.}) \pm1.0({\rm syst.})$ & \protect\cite{Alamos1}
     &     &     &     &     \\
     & $ 8.9\pm0.6({\rm stat.}) \pm0.75({\rm syst.})$ & \protect\cite{Drexlin}
     & 8.9 & 8.9 & 9.3 & 9.3 \\
     & $ 9.1\pm0.4({\rm stat.}) \pm0.9({\rm syst.})$ & \protect\cite{LSND}
     &     &     &     &     \\
   $^{12}$C($\nu_\mu,\mu^-$)$^{12}$N$_{\rm g.s.}$  
     & $ 66 \pm 10({\rm stat.}) \pm 10({\rm syst.})$ & \protect\cite{LSND}   
     &  68{ } &  73{ } &  63{ } &  63{ } \\ \hline
   ${}^{12}{\rm C} (\nu,\nu^\prime) {}^{12}{\rm C}^*(15.11)$ 
     & $10.4\pm1.0({\rm stat.}) \pm0.9({\rm syst.})$ & \protect\cite{neutral}
     & 10.5 & 10.5 & 10.5 & 10.6 \\
   ${}^{12}{\rm C} (\nu_\mu,\nu_\mu^\prime) {}^{12}{\rm C}^*(15.11)$ 
     & $ 3.2 \pm 0.5 ({\rm stat.}) \pm0.4 ({\rm syst.})$ 
     & \protect\cite{Karmen3}
     & 2.8 & 2.7 & 2.8 & 2.8 \\ 
\end{tabular}
\end{center}
\end{table}

\begin{table}
\begin{center}
\caption
{The inclusive muon capture rate $\omega$, 
(in $10^3$ s$^{-1}$)
and the cross sections $\sigma$ for the
$^{12}$C($\nu_e,e^-$)$^{12}$N$^*$ (in units of $10^{-42}$ cm$^2$) and
the total (inclusive + exclusive) cross section for the 
$^{12}$C($\nu_\mu,\mu^-$)$^{12}$N (in $10^{-40}$ cm$^2$) reactions..
The results of our improved continuum RPA calculation 
($(\omega/\sigma)^{\rm new}$) are compared to the data and the previous 
($(\omega/\sigma)^{\rm old}$) 
continuum RPA calculation without partial $p_{1/2}$
subshell occupation.}
\begin{tabular}{|c|cr|rr|rr|} \hline
   process & data & Ref. & $(\omega/\sigma)_{\rm BP}^{\rm new}$ 
                         & $(\omega/\sigma)_{\rm LM}^{\rm new}$ 
                         & $(\omega/\sigma)_{\rm BP}^{\rm old}$ 
                         & $(\omega/\sigma)_{\rm LM}^{\rm old}$ \\ \hline
   $^{12}$C($\mu^-,\nu_\mu$)$^{12}$B$^*$ & $32.8 \pm 0.8$ & \cite{bike}
                         & 32.7 & 31.3 & 34.2 & 33.3 \\
   $^{12}$C($\nu_e,e^-$)$^{12}$N$^*$ & $ 5.1 \pm 0.6 \pm 0.5$ & 
                         \cite{Kleinfeller}
                         & 5.4  & 5.6 & 6.3 & 5.9  \\
                         & $5.7 \pm 0.6 \pm 0.6$ & \cite{Imlay}   
                         &      &     &     &      \\ 
   $^{12}$C($\nu_\mu,\mu^-$)$^{12}$N & $12.4 \pm 0.3 \pm 1.8$ & \cite{Imlay2}  
                         & 17.8 & 17.5 & 19.3 & 20.3 \\
\end{tabular}
\end{center}
\end{table}

\begin{table}
\begin{center}
\caption
{The full strength within the nuclear $p$ shell evaluated for the
operators in column 1. The `SM' column is the exact shell model result
calculated with the Cohen-Kurath interaction. The column named `naive'
corresponds to the $(p_{3/2})^8$ configuration. In the last column the
strength for transitions with $2 \hbar \omega$ is shown for comparison.}
\begin{tabular}{|c|c|c|c|} \hline
Operator & SM & naive & $2 \hbar \omega$ \\
\hline
Gamow-Teller ($\sigma \tau$) & 1.51 & 8.00 &  0.00 \\
$r^2 Y_2$                    & 1.37 & 1.98 &  9.95 \\ 
$r^2 (Y_2 \sigma)^{I=1}$     & 0.11 & 0.08 &   -   \\
$r^2 (Y_2 \sigma)^{I=2}$     & 0.33 & 0.75 &   -   \\
$r^2 (Y_2 \sigma)^{I=3}$     & 0.20 & 0.00 &   -   \\
$\sum_\lambda r^2 (Y_2 \sigma)^\lambda$     & 0.64 & 0.83 &  7.47 \\
\end{tabular}
\end{center}
\end{table}

\begin{table}
\caption{Total $^{12}$C($\nu_e,e^-$)$^{12}$N, 
               $^{12}$C(${\bar \nu_e},e^+$)$^{12}$B, 
               and exclusive cross sections to the $^{12}$N and $^{12}$B
               ground states for a mesh of neutrino energies $E_{\nu_e}$. 
               The cross sections have been calculated with the BP-interaction 
               and are given in units of $10^{-42}$ cm$^2$.
               Energies are in MeV, exponents are given in parentheses.}
 \begin{center}
  \begin{tabular}{|r|l|l|l|l|} \hline
    $E_{\nu_e}$  & 
    $^{12}$C($\nu_e,e^-$)$^{12}$N & 
    $^{12}$C($\nu_e,e^-$)$^{12}$N$_{gs}$ & 
    $^{12}$C(${\bar \nu_e},e^+$)$^{12}$B &
    $^{12}$C(${\bar \nu_e},e^+$)$^{12}$B$_{gs}$   \\ \hline
     20 & 2.85 (-1) &  2.84 (-1) & 8.00 (-1) &  7.90 (-1) \\
     30 & 5.65      &  4.90      & 6.05      &  5.01      \\
     40 & 2.23 (+1) &  1.46 (+1) & 1.84 (+1) &  1.17 (+1) \\
     50 & 5.99 (+1) &  2.79 (+1) & 4.14 (+1) &  1.95 (+1) \\
     60 & 1.31 (+2) &  4.31 (+1) & 7.80 (+1) &  2.72 (+1) \\ 
     70 & 2.47 (+2) &  5.82 (+1) & 1.30 (+2) &  3.41 (+1) \\
     80 & 4.21 (+2) &  7.16 (+1) & 1.99 (+2) &  3.97 (+1) \\ 
     90 & 6.63 (+2) &  8.25 (+1) & 2.84 (+2) &  4.39 (+1) \\
    100 & 9.79 (+2) &  9.03 (+1) & 3.86 (+2) &  4.69 (+1) \\ 
    110 & 1.37 (+3) &  9.49 (+1) & 5.01 (+2) &  4.90 (+1) \\
    120 & 1.85 (+3) &  9.70 (+1) & 6.29 (+2) &  5.04 (+1) \\ 
    130 & 2.41 (+3) &  9.70 (+1) & 7.68 (+2) &  5.14 (+1) \\
    140 & 3.06 (+3) &  9.58 (+1) & 9.17 (+2) &  5.22 (+1) \\ 
    150 & 3.79 (+3) &  9.37 (+1) & 1.08 (+3) &  5.29 (+1) \\
    160 & 4.61 (+3) &  9.15 (+1) & 1.24 (+3) &  5.35 (+1) \\ 
    170 & 5.51 (+3) &  8.94 (+1) & 1.42 (+3) &  5.41 (+1) \\
    180 & 6.49 (+3) &  8.77 (+1) & 1.59 (+3) &  5.47 (+1) \\ 
    190 & 7.54 (+3) &  8.63 (+1) & 1.78 (+3) &  5.52 (+1) \\
    200 & 8.66 (+3) &  8.51 (+1) & 1.96 (+3) &  5.58 (+1) \\ 
    210 & 9.84 (+3) &  8.42 (+1) & 2.15 (+3) &  5.63 (+1) \\
    220 & 1.11 (+4) &  8.35 (+1) & 2.34 (+3) &  5.68 (+1) \\ 
    230 & 1.23 (+4) &  8.28 (+1) & 2.53 (+3) &  5.72 (+1) \\ 
    240 & 1.36 (+4) &  8.22 (+1) & 2.72 (+3) &  5.77 (+1) \\
    250 & 1.50 (+4) &  8.17 (+1) & 2.91 (+3) &  5.81 (+1) \\ 
    260 & 1.63 (+4) &  8.11 (+1) & 3.11 (+3) &  5.85 (+1) \\
    270 & 1.76 (+4) &  8.06 (+1) & 3.30 (+3) &  5.89 (+1) \\
    280 & 1.90 (+4) &  8.01 (+1) & 3.50 (+3) &  5.93 (+1) \\
    290 & 2.03 (+4) &  7.96 (+1) & 3.70 (+3) &  5.97 (+1) \\
    300 & 2.16 (+4) &  7.91 (+1) & 3.90 (+3) &  6.00 (+1) \\ \hline
\end{tabular}
\end{center}
\end{table}

\end{document}